\documentclass[twocolumn,showpacs,preprintnumbers,amsmath,amssymb]{revtex4}
\usepackage{color}
\def\blue#1{\textcolor{blue}{#1}}

\begin{document}

\title{Quantum tests for the linearity and permutation invariance of Boolean functions}
\author{ Mark Hillery$^1$ and Erika Andersson$^2$}
\affiliation{
$^{1}$Department of Physics, Hunter College of the City University of New York, 695 Park Avenue, New York, NY 10021 USA \\
$^2$SUPA, School of Engineering and Physical Sciences, Heriot-Watt University, Edinburgh EH14 4AS, UK}

\begin{abstract}
The goal in function property testing is to determine whether a black-box Boolean function has a certain property
or is $\epsilon$-far from having that property.  The performance of the algorithm is judged by how many calls
need to be made to the black box in order to determine, with high probability, which of the two alternatives is the
case.  Here we present two quantum algorithms, the first to determine whether the function is linear and the 
second to determine whether it is symmetric (invariant under permutations of the arguments).  Both require
order $\epsilon^{-2/3}$ calls to the oracle, which is better than known classical algorithms.  In addition, in the case
of linearity testing, if the function is linear, the quantum algorithm identifies which linear function it is.  The linearity
test combines the Bernstein-Vazirani algorithm and amplitude amplification, while the test to determine whether
a function is symmetric uses projective measurements and amplitude amplification.
\end{abstract}

\maketitle

\section{Introduction}
One of the first quantum algorithms to be discovered was the Bernstein-Vazirani algorithm \cite{bernstein}.  This
algorithm allows one to identify an unknown linear Boolean function with only one call to the oracle, or black
box, that evaluates that function.  Classically, if the inputs to the function are $n$-bit strings, $n$ calls would be
required.  A subsequent quantum algorithm, the Grover algorithm, also identifies an unknown Boolean
function \cite{grover}.   In the simplest  case of its use, the set of functions being considered consists of those 
functions whose inputs  are $n$-bit 
strings and whose outputs are $0$ on all of the strings except one.  The Grover algorithm can find to which
Boolean function the oracle corresponds (or which string gives the output $1$) with order $2^{n/2}$ calls to the 
oracle rather than the order $2^{n}$ that would be required classically.  
Here we would like to consider two additional
quantum algorithms that apply to Boolean functions.  Both make use of a generalization of the Grover algorithm
known as amplitude amplification \cite{brassard} and one also makes use of the Bernstein-Vazirani algorithm.
Both determine whether an unknown Boolean function has a particular property or is far from having that
property.  Problems of this type fall into the area of function property testing.  The first algorithm presented here
will test whether a function is linear, and the second will test whether it is symmetric.  Both perform better than
existing classical algorithms.

Now let us discuss what our algorithms do in somewhat more detail.
Function property testing is an area of computer science that finds algorithms to determine whether a black-box
Boolean function has a certain property or is far from having that property.  A Boolean function is one whose 
inputs are $n$-bit strings, $x_{1}x_{2}\ldots x_{n}$, and whose output is either $0$ or $1$.  One of the 
properties one can test for is linearity; a Boolean function is linear if and only if it can be expressed as
\begin{equation}
f(x_{1},x_{2},\ldots x_{n}) = a_{1}x_{1} + a_{2}x_{2} + \ldots a_{n}x_{n}  ,
\end{equation}
where $a_{j}$ is either $0$ or $1$, and all operations are modulo $2$.  We can express the above equation
as $f(x)=a\cdot x$, where $x$ and $a$ are $n$-bit strings, and the dot product of two strings is defined as above.
An equivalent definition of linearity is that a Boolean function is linear if and only if it satisfies $f(x+y)=f(x)+f(y)$, where $x$ and $y$ are $n$-bit strings, and $x+y$ is the $n$-bit string whose $j^{\rm th}$ element is $x_{j}+y_{j}$.  There is a classical test for linearity, known as the BLR (Blum, Luby, Rubinfeld) test \cite{blr}, and what we wish to do is to develop a quantum test  that requires fewer calls to the oracle.
A second property we shall test for is whether a function is symmetric.  A Boolean function is symmetric if it
is invariant under all permutations of its arguments.

Quantum property testing was first considered by Buhrman et al. \cite{buhrman}.  They found situations for which
there are quantum algorithms that are better than any classical algorithm, in terms of the number of calls to the oracle, and in some cases exponentially better.
Atici and Serviedo discuss a quantum algorithm for testing whether a Boolean function is a $k$-junta \cite{atici}.  A Boolean function is a $k$-junta if it depends on
only $k$ of the $n$ variables.  One can also devise quantum algorithms to identify which input variables a Boolean function depends on, and for learning the form of quadratic and cubic Boolean functions~\cite{dominik}. R\"otteler~\cite{rotteler} has also discussed quantum algorithms to identify quadratic Boolean functions.  More recently, Montanaro and Osborne defined quantum Boolean functions and
developed several property testing algorithms for them \cite{montanaro}.  The Boolean functions we consider in
this paper will be strictly classical.  Finally, Ambainis, Childs and Liu have developed algorithms for testing the
properties of graphs \cite{ambainis}.

We will begin by discussing some of the basic ideas of function testing and then go on to present the classical BLR algorithm for linearity testing.  Because our algorithm is a combination of two existing quantum algorithms, the Bernstein-Vazirani algorithm and the Grover algorithm, we will review some features of both of these.  We will then present our quantum algorithm for linearity testing.  Next we shall discuss the classical algorithm for testing whether a Boolean function is symmetric, and then go on to present a quantum algorithm that does so with
fewer oracle calls.

\section{Function testing and the BLR test}
As was mentioned in the Introduction, a Boolean function maps $n$-bit strings to $\{ 0,1\}$.  We say that two 
Boolean functions, $f$ and $g$, are $\epsilon$-close if they agree on at least a $(1-\epsilon )$ fraction of their inputs.  Another way of saying this is to define a distance between $f$ and $g$ as
\begin{equation}
d(f,g) = \frac{1}{N}\sum_{x} |f(x) - g(x)| ,
\end{equation}
were $N=2^{n}$, which is just the fraction of strings on which $f$ and $g$ disagree.  So, $f$ and $g$ are
$\epsilon$-close if and only if $d(f,g) \leq \epsilon$.
If they are not $\epsilon$-close, then they are $\epsilon$-far.  We say that a function, $f$, is $\epsilon$-close 
to having a particular property, if there is a function, $g$, that has that property that is $\epsilon$-close to $f$.
If there is no such function, then $f$ is said to be $\epsilon$-far from having that property.  For a discussion
of these definitions, as well as a very readable discussion of function testing in general, see \cite{odonnell}.

In the quantum case, it is also useful to think of Boolean functions as vectors in a Hilbert space.  
The space is just $\mathcal{H}=\mathcal{H}_{2}^{\otimes n}$, the space of $n$ qubits, where $\mathcal{H}_{2}$ 
is  the two-dimensional single-qubit space.  For the Boolean function $f(x)$, define the vector
\begin{equation}
|v_{f}\rangle = \frac{1}{\sqrt{N}} \sum_{x} (-1)^{f(x)} |x\rangle ,
\end{equation}
where $|x\rangle$ is a state in the computational basis, and, as before, $N=2^n$.  This vector is generated in 
a very natural way by the quantum oracle $U_{f}$ that evaluates $f(x)$.  The operation $U_{f}$ is called an 
$f$-controlled-NOT  gate, and it acts as
\begin{equation}
U_{f}|x\rangle |b\rangle = |x\rangle |b+f(x)\rangle ,
\end{equation}
where $|b\rangle$ is a single-qubit state, with $b=0,1$, and the addition is modulo $2$.  
If $U_{f}$ is applied to the state $|x\rangle |-\rangle$, where $|-\rangle = (|0\rangle -|1\rangle )/\sqrt{2}$, the result is $(-1)^{f(x)} |x\rangle |-\rangle$, so that if $U_{f}$ is applied to $(1/\sqrt{N})\sum_{x} |x\rangle |-\rangle$, the result is $|v_{f}\rangle |-\rangle$.  If two functions $f$ and $g$ 
are $\epsilon$-close, then 
\begin{equation}
\langle v_{f}|v_{g}\rangle = [1-d(f,g)] - d(f,g) \geq 1-2\epsilon  .
\end{equation}

The vectors corresponding to linear Boolean functions form an orthonormal set, and they span $\mathcal{H}$,
so they constitute an orthonormal basis.   The orthonormality follows from the relation
\begin{equation}
\label{deltastring}
\frac{1}{N} \sum_{x\in \{ 0,1\}^{n}  } (-1)^{x\cdot y} = \delta_{y,0}  ,
\end{equation}
where $x$ and $y$ are $n$-bit strings.  Because these vectors are orthonormal, they are perfectly distinguishable,
and this is, in fact, the basis of the Bernstein-Vazirani algorithm \cite{bernstein}.  The problem that this algorithm solves is the following.  One is given a black box that evaluates some linear Boolean function, and the task is to determine which Boolean function it evaluates.  The Bernstein-Vazirani algorithm accomplishes this with one query to the black box.  The fact that the black boxes corresponding to different linear Boolean functions can be used to produce orthogonal vectors implies that with a single measurement we can perfectly determine which function we have.

This is actually accomplished by using a circuit consisting of Hadamard gates and an 
$f$-controlled-NOT gate.  If we apply $n$ Hadamard gates, one to each qubit, in the state
$|x\rangle$, we obtain
\begin{equation}
H^{\otimes n}|x\rangle = \frac{1}{\sqrt{N}}\sum_{y\in \{ 0,1\}^{n} } (-1)^{x\cdot y} |y\rangle ,
\end{equation}
where, as before, we have set $N=2^{n}$.  Now, the input state to our circuit is the $(n+1)$
qubit state 
\begin{equation}
|\Psi_{in}\rangle  =\frac{1}{\sqrt{2}}|00\ldots 0\rangle (|0\rangle - |1\rangle ) .
\end{equation}
We first apply $n$ Hadamard gates, one to each of the first $n$ qubits, and then the 
$f$-controlled-NOT gate, giving us
\begin{equation}
|\Psi_{in}\rangle \rightarrow \frac{1}{\sqrt{2N}} \sum_{x\in \{ 0,1\}^{n}  } (-1)^{f(x)}|x\rangle
(|0\rangle - |1\rangle ).
\end{equation}
At this point, since $f(x)$ is linear, let us set it equal to $f(x)=a\cdot x$, where now the object has become to
determine the $n$-bit string $a$. Next,  we again apply $n$ Hadamard gates to the first $n$ qubits yielding 
\begin{equation}
|\Psi_{out}\rangle = \frac{1}{N\sqrt{2}}  \sum_{x\in \{ 0,1\}^{n}  } \sum_{y\in \{ 0,1\}^{n}  }
(-1)^{x\cdot (a+y)}|y\rangle (|0\rangle - |1\rangle ) .
\end{equation}
Discarding the last qubit (it is not entangled with the others), and taking note of Eq.\ (\ref{deltastring}), we
see that we are left with the output state $|a\rangle$, which we can just measure in the computational
basis to find the $n$-bit string $a$.  Therefore, we find out what the function 
is with only one application of the $f$-controlled-NOT gate.  Classically, we would need to 
evaluate the function $n$ times to find $a$.

If $f(x)$ is not linear, the output vector can be expressed as
\begin{equation}
|\Psi_{out}\rangle = \frac{1}{\sqrt{2}}   \sum_{y\in \{ 0,1\}^{n}  } \langle v_{f}|v_{y\cdot x}\rangle |y\rangle 
(|0\rangle - |1\rangle ) ,
\end{equation}
where $|v_{y\cdot x}\rangle$ is the vector corresponding to the function $g(x)=y\cdot x$.  In this case, if we
measure $|\Psi_{out}\rangle$ in the computational basis, the probability of obtaining the $n$-bit string $z$ is
$| \langle v_{f}|v_{z\cdot x}\rangle|^{2} = | 1-2d(f,z\cdot x)|^{2}$~\cite{dominik}.

In the Grover algorithm, we successively apply what is known as the Grover operator to an initial state, and this 
has the effect of rotating that state into the desired state.  In the usual case, when one wishes to find for which
input $x^{\prime}$ the unknown function is one, i.e. $f(x^{\prime})=1$, and it holds that $f(x)=0$ for $x\neq x^{\prime}$, the 
initial state is $(1/\sqrt{N})\sum_{x}|x\rangle$ and the desired state is $|x^{\prime}\rangle$.  All of the action in
the Grover algorithm takes place in the two-dimensional real vector space spanned by these two vectors.  In
general, if the two-dimensional real space is spanned by the vectors $|v_{1}\rangle$ and $|v_{2}\rangle$, the
Grover operator will be of the form \cite{jozsa}
\begin{equation}
G= (I-2|v_{2}^{\perp}\rangle\langle v_{2}^{\perp}|) (I- 2|v_{1}^{\perp}\rangle\langle v_{1}^{\perp}|) ,
\end{equation}
where $|v_{1}^{\perp}\rangle$ is orthogonal to $|v_{1}\rangle$ and $|v_{2}^{\perp}\rangle$ is orthogonal to
$|v_{2}\rangle$.
This is a product of two reflections in the two-dimensional real space ($|v_{1}\rangle$ and $|v_{2}\rangle$ can be visualized as two vectors in the Euclidean plane), one about the line containing $|v_{1}\rangle$ and one about the line containing $|v_{2}\rangle$.  It is a theorem in plane geometry that the product of two reflections is a rotation
by twice the angle between the lines, in this case twice the angle between $|v_{1}\rangle$ and $|v_{2}\rangle$.  We will be using the Grover algorithm to rotate an initial vector in the direction of a component of the function we are
testing that does not have the desired property, e.g. linearity, should such a component exist.

Now let us look at a classical test for deciding whether a Boolean function is linear or $\epsilon$-far from being
linear, the BLR test \cite{blr,odonnell}.  The function is promised to be to belong to one of these two alternatives, that is, it is guaranteed to be either linear or $\epsilon$-far from linear.  A single instance of the BLR procedure goes as follows:
\begin{itemize}
\item Pick two $n$-bit strings $x$ and $y$ independently and uniformly at random from $\{ 0,1\}^n$.
\item Set $z=x+y$.
\item Query $f$ on $x$, $y$, and $z$.
\item Accept if $f(z)=f(x)+f(y)$.
\end{itemize}
This test has the following properties \cite{odonnell}: 
\begin{itemize}
\item If a function is linear, the probability the test accepts is one.
\item If a function is $\epsilon$-far from linear, the probability the test accepts is less than $1-\epsilon$.
\end{itemize}
In order to decide whether a function is linear, we 
run the test of the order $1/\epsilon$ times and overall accept if each individual test accepts.  
Note that the probability that a function that is $\epsilon$-far from linear will be accepted on each
of $m$ runs, $p_{m}$, is
\begin{equation}
p_{m} \leq (1-\epsilon )^{m} = e^{m\ln (1-\epsilon)} \leq e^{-m\epsilon} ,
\end{equation}
so by choosing $m$ of order $\epsilon^{-1}$ we can make the probability of accepting a function that is indeed 
$\epsilon$-far from linear quite small.  For example, if we would like to make this probability less than $1/3$,
we can choose $m>(\ln 3)/\epsilon$.

\section{Quantum algorithm}
We will now describe a quantum algorithm to determine whether a function is linear or $\epsilon$-far from
linear.  Again, our function is promised to be either linear or $\epsilon$-far from linear.
If the function is linear it will definitely give ``yes."  If it is  $\epsilon$-far from linear, 
it will say ``no" with probability greater than 2/3.  It requires of the order of $\epsilon^{-2/3}$ oracle calls, and has the additional property that if the function is linear, it tells us which linear function it is.  Schematically, the algorithm
is as follows.
\begin{itemize}
\item First run the  Bernstein-Vazirani algorithm on the function of the order of $\epsilon^{-2/3}$ times.  If one gets the same result every time (the same linear function) one then proceeds to the next step.  If not, the function is declared to be $\epsilon$-far from linear.
\item If the linear function we obtained in the first step is $g(x)$, we use the $f$-controlled-NOT gate to generate 
the state $(1/\sqrt{N}) \sum_{x} |x\rangle |f(x)+g(x)\rangle$ (the addition is modulo $2$) and measure the last
qubit in the computational basis.  If we obtain $0$ we proceed to the next step, if not, the function is declared
to be $\epsilon$-far from linear.
\item We make use of the candidate linear function to construct a Grover-like algorithm that amplifies the nonlinear part, if there is one, of the function we are testing.  This algorithm is then run of the order of $\epsilon^{-1/3}$ times, each time for of the order of $\epsilon^{-1/3}$ steps. After each run of the Grover algorithm, the system is measured to see if it is  still in the state corresponding to the candidate linear function.  If the function passes this test, it is declared to be linear, if not, it is declared to be $\epsilon$-far from linear.
\end{itemize}

A linear function will be declared to be linear by this test.  Now let us see what happens if the function 
is $\epsilon$-far from being linear.  Suppose we have a function $f$ that is $\epsilon$-far from 
being linear, and we have run the Bernstein-Vazirani algorithm once and obtained the linear function $g$
as our result.  Because $g$ is linear, we have $\langle v_{f}|v_{g}\rangle < 1-2\epsilon$.  Now
let us consider the rest of the first part of the test.  Let $a= \langle v_{f}|v_{g}\rangle < 1-2\epsilon$, where $g$ is
linear.  Then we can write
\begin{equation}
|v_{f}\rangle = a|v_{g}\rangle + |v_{g}^{\perp}\rangle  ,
\end{equation}
where $\langle v_{g}|v_{g}^{\perp}\rangle = 0$ and 
\begin{equation}
\| v_{g}^{\perp} \| = (1-a^{2})^{1/2} > 2\sqrt{\epsilon} (1-\epsilon )^{1/2} .
\end{equation}
Note that when $|v_{f}\rangle$ and $|v_{g}\rangle$ are expanded in the computational basis, the
resulting expansion coefficients are real.  This implies that $\langle v_{f}|v_{g}\rangle$ and $\langle v_{f}
|v_{g}^{\perp}\rangle$ are real.  We will split our analysis into three parts, $|a|<1-\epsilon^{2/3}$, 
$-1\leq a \leq -1+\epsilon^{2/3}$, 
and $1-\epsilon^{2/3} \leq a < 1-2\epsilon$.  These parts correspond to the three parts of our algorithm.
First consider $|a|<1-\epsilon^{2/3} = a_{0}$.  Suppose we run Bernstein-Vazirani $m$ more times 
and get $g$ each time.  The probability of this happening, $p_{g}(m)$, is
\begin{equation}
p_{g}(m) \leq | a_{0}|^{2m} = e^{2m\ln (1-\epsilon^{2/3})} \leq e^{-2m\epsilon^{2/3}} ,
\end{equation}
for $\epsilon < 1$.  Therefore, for $m$ of order $\epsilon^{-2/3}$, we
can make this probability small, in particular, it will be less than $1/3$ if $m>\ln 3/(2\epsilon^{2/3})$.

Let us now consider the case $-1\leq a \leq -1+\epsilon^{2/3}$.  It is necessary to single out this case, because
the Bernstein-Vazirani algorithm will return the same linear function $g(x)$ for two different inputs, $g(x)$
and $\bar{g}(x)=1+g(x)$.  Note that $\langle v_{\bar{g}}|v_{g}\rangle = -1$.  So, if $f$ passes the first step of the
algorithm, we need to ensure that $\langle v_{f} |v_{g}\rangle$ is close to $1$ and not close to $-1$.  In order to
do this, as stated above, we use the $f$-controlled-NOT gate to produce the state $(1/\sqrt{N}) \sum_{x} |x\rangle 
|f(x)+g(x)\rangle$ and measure the last qubit.  The probability of obtaining $0$ is $1-d(f,g)=(1+\langle v_{f}
|v_{g}\rangle )/2$.  For $-1\leq a \leq -1+\epsilon^{2/3}$, this is less than $(1/2)\epsilon^{2/3}$.  For $\epsilon <1/8$,
this will be less than $1/3$.

Now we will consider the case $1-\epsilon^{2/3} \leq a < 1-2\epsilon$, and this is where amplitude amplification
comes in.  We now assume that we have performed the Bernstein-Vazirani part of the algorithm and gotten the linear function $g$ each time.  Define the operator
\begin{equation}
M=(I-2|v_{f}\rangle\langle v_{f}|)(2|v_{g}\rangle\langle v_{g}| -I) ,
\end{equation}
and note that it can be realized with two applications or the oracle (the oracle is used to generate  
the operator $I-2|v_f\rangle\langle v_f|$).  In more detail, we have that 
\begin{eqnarray}
&&(I-2|v_{f}\rangle\langle v_{f}|)\otimes |-\rangle\langle -| \nonumber \\
&&= U_{f}H^{\otimes n}
[I- (|0\rangle\langle 0|)^{\otimes n} ] H^{\otimes n} \otimes |-\rangle\langle -| U_{f} .  \label{eq:Mconstr}
\end{eqnarray}
If the operation in Eq. \eqref{eq:Mconstr} is applied to a register of $n$ qubits plus an auxiliary qubit in the state 
$|-\rangle$, then the effect is to realize the operation $I-2|v_{f}\rangle\langle v_{f}|$ on the register. The auxiliary qubit can, as usual, be ignored after the operation, because it is not entangled with the rest of the state.  
It is straightforward to construct the operator $I-2|v_{g}\rangle\langle v_{g}|$ since we know $g$ explicitly.  The operator $M$ will rotate $|v_{f}\rangle$ toward $|v_{g}^{\perp}\rangle$.  If $|v_{g}^{\perp}\rangle =0$, as would be the case if $f$ is linear, then $M$ will simply have the effect of multiplying $|v_{g}\rangle$ by $-1$.  Therefore, we can see whether $|v_{f}\rangle$ has a component orthogonal to $|v_{g}\rangle$ by applying $M$ a number of times and measuring to see whether the resulting vector is still in the same direction as $|v_{g}\rangle$.

The operator $M$ acts in the two-dimensional real vector space spanned by 
$|v_{g}\rangle$ and $|v_{g}^{\perp}\rangle$.  Defining $|\tilde{v}_{g}\rangle = (1/\| v_{g}^{\perp} \| )
|v_{g}^{\perp} \rangle$, we can express $M$ in the basis $\{ v_{g}, \tilde{v}_{g} \}$ as
\begin{equation}
M=\left( \begin{array}{cc} 1-2a^{2} & 2a(1-a^{2})^{1/2} \\ -2a(1-a^{2})^{1/2} & 1-2a^{2} \end{array}
\right) .
\end{equation}
The eigenvalues and eigenvectors are
\begin{eqnarray}
\lambda_{\blue{1}} = 1-2a^{2} + 2ia(1-a^{2})^{1/2}; && |\eta_{1}\rangle = \frac{1}{\sqrt{2}}
\left( \begin{array}{c} 1 \\ i \end{array} \right)   \\
\lambda_{\blue{2}} = 1-2a^{2} - 2ia(1-a^{2})^{1/2}; && |\eta_{2}\rangle = \frac{1}{\sqrt{2}}
\left( \begin{array}{c} 1 \\ -i \end{array} \right).\nonumber
\end{eqnarray}
Defining 
\begin{equation}
e^{i\theta} = 1-2a^{2} + 2ia(1-a^{2})^{1/2} ,
\end{equation}
which implies that $\cos\theta = 1-2a^{2}$ and $\sin\theta = 2a(1-a^{2})^{1/2}$, we find that
\begin{equation}
M^{n}|v_{f}\rangle = \left( \begin{array}{c} a\cos n\theta + (1-a^{2})^{1/2}\sin n\theta \\
-a\sin n\theta + (1-a^{2})^{1/2}\cos n\theta \end{array} \right) .
\end{equation}
After the $n$ applications of $M$ we measure the projection $P_{g}=|v_{g}\rangle\langle v_{g}|$. 
The probability that we obtain $0$, which indicates that the function is not linear, is
\begin{eqnarray}
q(a,n) & = & [ -a\sin n\theta + (1-a^{2})^{1/2}\cos n\theta ]^{2}  \nonumber  \\
 & = & \frac{1}{2} + \frac{1}{2} [(1-2a^{2})\cos (2n\theta ) \nonumber \\
 & & - 2a (1-a^{2})^{1/2} \sin (2n\theta )] \nonumber \\
 & = & \frac{1}{2} \{ 1 + \cos [(2n+1)\theta ] \}  .
\end{eqnarray}
If $\epsilon$ is small, $\theta$ will be close to, but less than, $\pi$, and so  we can express it as
$\theta = \pi - \delta\theta$, where $\delta\theta$ is small and positive.  This gives us for $q(a,n)$
\begin{equation}
q(a,n) = \frac{1}{2} \{ 1- \cos [(2n+1)\delta\theta ] \} .
\end{equation}
Making use of the bound
on $a$, $a_{0} \leq a< 1-2\epsilon$, which implies that $1-a_{0}^{2} \geq 1-a^{2} > 4\epsilon 
(1-\epsilon )$, we have that
\begin{equation}
\label{sin-ineq}
2(1-2\epsilon )(1-a_{0}^{2})^{1/2} \geq 2a(1-a^{2})^{1/2} > 4\sqrt{\epsilon} (1-\epsilon)^{1/2} a_{0} .
\end{equation}

Let us now do a rough calculation to give the basic idea of this part of the algorithm.  A more detailed calculation
will follow.  Now, the quantity in the middle of the above inequality is just $\sin\theta = \sin (\delta\theta )$. For
$\epsilon$ sufficiently small, we see that, using $\sin\delta\theta \simeq \delta\theta$,  
\begin{equation}
2\sqrt{2}\epsilon^{1/3} \geq \delta \theta \geq 4\sqrt{\epsilon} .
\end{equation}
We now choose $n$ so that $(2n+1)2\sqrt{2}\epsilon^{1/3} =\pi$.  This guarantees that if $\delta \theta$ is at the
top of its range, we will have $q(n,a)=1$, and we will find after one measurement of $P_{g}$ that the function is
not linear.  Note that in this case, this part of the algorithm makes of order $\epsilon^{-1/3}$ function calls.  Now,
for this value of $n$, the worst case is if $\delta\theta$ is at the bottom of its range.  We then have that $(2n+1)
\delta\theta \sim \epsilon^{1/6}$.  Assuming $\epsilon$ is sufficiently small so that $(2n+1)\delta\theta$ is much
less than one, we have that $q(a,n) \simeq (1/4)[(2n+1)\delta\theta ]^{2}$, so that
\begin{equation}
q(a,n) \simeq \frac{\pi^{2}}{2} \epsilon^{1/3} .
\end{equation}
This probability is small, but if we repeat this process, run Grover for $\epsilon^{-1/3}$ steps and then measure 
$P_{g}$, $\epsilon^{-1/3}$ times, the probability that we will get at least one measurement result of $0$ if the function is $\epsilon$-far from linear will be of order one.  In this case we make of order $\epsilon^{-2/3}$ function calls.  So, the total number of function calls is of order $\epsilon^{-2/3}$ for the first part of the algorithm, and of order $\epsilon^{-2/3}$ for the second part (using the worst case number), for a total of order $\epsilon^{-2/3}$ calls
for the entire algorithm.

Now let us do this in more detail.  We shall assume for now that $\epsilon <1/8$, as we have done so far, 
but we will find, in the course of our analysis, that this is not sufficient, and that a smaller range will be required.
Going back to Eq.\ (\ref{sin-ineq}), and using the fact that
for $0\leq \theta \leq \pi /2$, if $k_{1} \geq \sin\theta \geq k_{2}$, then $(\pi /2)k_{1} \geq \theta \geq k_{2}$, and
assuming that $\epsilon <1/8$, we find that $\pi\sqrt{2}\epsilon^{1/3} > \delta\theta > (5/2)\epsilon^{1/2}$.  This follows from 
\begin{eqnarray}
\frac{\pi}{2} k_{1} & = & \pi \epsilon^{1/3} (1-2\epsilon )(2-\epsilon^{2/3})^{1/2} < \pi \sqrt{2}\epsilon^{1/3} 
\nonumber \\
k_{2} & = & 4\sqrt{\epsilon} (1-\epsilon )^{1/2} (1-\epsilon^{2/3}) >\frac{5}{2} \epsilon^{1/2} ,
\end{eqnarray}
where the numbers in the expressions at the right were found by making use of the condition  $\epsilon <1/8$.
We are now going to run Grover's algorithm for $n$ steps, where
\begin{equation}
\label{nvalue}
(2n+1)\sqrt{2} \epsilon^{1/3} = 1 ,
\end{equation}
so that $n$ is of order $\epsilon^{-1/3}$.  Now, $2n+1$ must be an odd integer, and the above equation will, in
general, not give us this result.  So, we choose $n$ so that $2n+1$ is the closest odd integer to 
$1/(\sqrt{2} \epsilon^{1/3})$.   
If $\delta\theta$ is at the top of its allowed range, then this will result in $q(a,n)$ of order $1$, and when we
do our measurement we will find that the function was not linear.  Now we have to see what happens
if $\delta\theta$ is at the bottom of its allowed range.  This should be the worst case, since $n$ has been tuned for the top of the allowed range.  Using the fact that for $0\leq \phi \leq \pi /2$, we have that
$\phi \geq \sin\phi \geq (2/\pi )\phi$, we find that
\begin{equation}
\label{estimate}
\frac{1}{\pi} \phi^{2} \leq 1-\cos\phi = \int_{0}^{\phi} d\phi^{\prime}\sin\phi^{\prime} \leq \frac{1}{2} \phi^{2} .
\end{equation}
This implies that
\begin{equation}
\frac{1}{2\pi} [(2n+1)\delta\theta ]^{2} \leq q(a,n) \leq   \frac{1}{4} [(2n+1)\delta\theta ]^{2} .
\end{equation}
If we take $n$ directly from Eq. (\ref{nvalue}), ignoring for the moment that $2n+1$ must be an odd integer, this would give us that 
\begin{equation}
\frac{1}{2\pi}\left(\frac{25}{8}\right) \epsilon^{1/3} \leq q(a,n) \leq \frac{1}{4}\left(\frac{25}{8}\right) 
\epsilon^{1/3}  ,
\end{equation}
where we have set $\delta \theta = (5/2)\epsilon^{1/2}$.
This inequality, in fact, gives us the dominant behavior as $\epsilon \rightarrow 0$.  However, we do need to take
into account the fact that $2n+1$ must be an odd integer.  This implies that $(\sqrt{2} \epsilon^{1/3})^{-1} -1 \leq
 2n+1\leq (\sqrt{2} \epsilon^{1/3})^{-1}+1$, so that
 \begin{eqnarray}
 \frac{1}{2\pi}\left( \frac{5}{2\sqrt{2}}\epsilon^{1/6}-\frac{5}{2}\epsilon^{1/2} \right)^{2} \leq q(a,n) \nonumber \\
 \leq \frac{1}{4}\left( \frac{5}{2\sqrt{2}}\epsilon^{1/6}+\frac{5}{2}\epsilon^{1/2} \right)^{2} .
\end{eqnarray}
Thus, we see that $q(a,n)$ is of order $\epsilon^{1/3}$ with corrections of order $\epsilon^{2/3}$.  Now, in order
for the $\epsilon^{1/3}$ behavior to be dominant, we need $\epsilon$ to be sufficiently small so that the ratio of
the corrections, of order $\epsilon^{2/3}$, to $\epsilon^{1/3}$ be small, i.e. $\epsilon^{1/3}\ll 1$.  If we now
choose this ratio to be less than $1/10$, this implies that $\epsilon \leq 10^{-3}$.   

We shall henceforth assume that $\epsilon \leq 10^{-3}$.  This allows us to sharpen our bounds for $\delta\theta$.
We first note that in this case, from the upper bound in Eq.\ (\ref{sin-ineq}), $\sin\delta\theta < 2\sqrt{2}(0.1) <1/2$,
which implies that $\delta\theta < \pi /6$.  Now, for $0\leq \theta \leq \pi /6$ we have that if $k_{1}\geq \sin\theta
\geq k_{2}$, then $(\pi /3)k_{1} \geq \theta \geq k_{2}$.  This now gives us that
\begin{equation}
\frac{2\pi\sqrt{2}}{3} \epsilon^{1/3} \geq \delta \theta \geq (3.996)\epsilon^{1/2} ,
\end{equation}
a tighter bound than before.  We now choose $2n+1$ to be the closest odd integer to 
$[3/(2\sqrt{2})]\epsilon^{-1/3}$, which means that if $\delta\theta$ is a the top of its range, $q(n,a)$ will be close to
one, and if the function is $\epsilon$-far from linear, this will be detected (by obtaining the measurement result $0$ when $P_{g}$ is measured) after a small number of runs of the amplitude amplification algorithm.  Let us now 
see what happens in the worst case for this choice of $n$, i.e.\ when $\delta\theta$ is at the bottom of its range.  Setting $\alpha = [3(3.996)]/[2\sqrt{2}]$, we have that
\begin{equation}
\frac{1}{2\pi}[ \alpha \epsilon^{1/6}-4\epsilon^{1/2}]^{2} \leq q(n,a) \leq \frac{1}{4}[ \alpha \epsilon^{1/6}
+ 4\epsilon^{1/2}]^{2} .
\end{equation}
The terms proportional to $\sqrt{\epsilon}$ are the corrections due to the fact that $(2n+1)$ is an odd integer, and
it can be seen that the ratio of  $4\sqrt{\epsilon}$ to the dominant $\alpha \epsilon^{1/6}$ term is less than $1/10$
for $\epsilon = 10^{-3}$ and decreases as $\epsilon^{1/3}$ as $\epsilon \rightarrow 0$.  

From the inequality, we see that $q(a,n)\sim \epsilon^{1/3}$.
This probability is small, but if we repeat this procedure, run Grover for $n$ steps and measure, $r$
times, the probability that we never get $0$ when we measure $P_{g}$ $r$ times is
\begin{eqnarray}
[1-q(a,n)]^{r} &  \leq & \left[ 1- \frac{1}{2\pi}\alpha^{2} \epsilon^{1/3}\right]^{r} \nonumber \\
& \leq & \exp \left[- r \frac{1}{2\pi}\alpha^{2} \epsilon^{1/3}\right]   ,
\end{eqnarray}
where we have ignored the $ 4\epsilon^{1/2}$ corrections.  This 
can be made small if we choose $r$ of order $\epsilon^{-1/3}$.  In particular, if 
\begin{equation}
r>  \frac{2\pi}{\alpha^{2}} \epsilon^{-1/3}  \ln 3  ,
\end{equation} 
then the probability of never getting $0$ when we measure $P_{g}$ is less than $1/3$.

Summarizing, we found that if $f$ is $\epsilon$-far from linear, and $|a|<a_{0}$, we will find that it
is not linear with a probability of order one by running Bernstein-Vazirani order $\epsilon^{-2/3}$
times.  In the case that $a_{0}<a<1-2\epsilon$, assuming we get the same linear function every
time we run Bernstein-Vazirani, then by running Grover order $\epsilon^{-1/3}$ steps order
 $\epsilon^{-1/3}$ times, for a total of  order $\epsilon^{-2/3}$ function calls, we will with a probability
 or order one detect the fact that it is not linear.  In both cases the total number of function calls is of order
 $\epsilon^{-2/3}$.

\section{Testing permutation invariance}
We now want to present a variant of the algorithm in the previous section that can test whether 
a Boolean function is invariant under permutations of its arguments, or is $\epsilon$-far from having this property.
As was noted in the Introduction, a function that is invariant under all permutations of its arguments is called symmetric.  Another way of phrasing this is that we are testing whether a function depends only on the Hamming weight of  its arguments or is $\epsilon$-far from having this property.  The Hamming weight of
the sequence $x=x_{1}x_{2}\ldots x_{n}$ is just the number of ones in the sequence, so that if $f(x_{1},x_{2},
\ldots x_{n})$ depends only on the Hamming weight of its arguments, its value is determined only by how
many of the $x_{j}$, for $1\leq j \leq n$ are equal to one. 

There is a classical algorithm to test whether a Boolean function is symmetric or $\epsilon$-far from being symmetric \cite{majewski}.  Again, it should be emphasized that this is a promise problem, the function is
guaranteed to be one or the other.  The procedure is to randomly choose an $n$-bit input, $x$, that is not either all
zeroes or all ones and evaluate $f(x)$.  One then chooses an input $y\neq x$ that has the same Hamming weight as $x$.  Next, one checks and sees whether $f(x)=f(y)$, and, if so, outputs ``yes,'' otherwise one outputs ``no.''  This procedure is repeated a number of times proportional to $\epsilon^{-1}$, and if one obtains ``yes'' every time, the function is declared to be symmetric.  If a ``no'' is obtained at any step the function is declared to be not symmetric.  A symmetric function will always be accepted as symmetric by this algorithm, and a function that is  $\epsilon$-far from being symmetric will be rejected with high probability.

Now let us go to our quantum algorithm.  Here the procedure is different than in the classical case.  We note that 
if a Boolean function is symmetric, then the corresponding vector, 
$|v_{f}\rangle$\blue{,} must lie in the completely symmetric subspace of $\mathcal{H}$.  Let us call this subspace $S$ 
and the projection operator onto it $P_{S}$.  Therefore, we would like to test whether a function is invariant under permutations of its arguments by testing whether the corresponding vector $|v_{f}\rangle$ is in $S$.  In order
to do so, we do the following:
\begin{itemize}
\item Measure $P_{S}$ order $\epsilon^{-2/3}$ times.  If any of our measurements yield $0$, we say the
function is $\epsilon$-far from symmetric, and stop.  If we do not obtain $0$ for any of our measurements 
results, we proceed to the next step.
\item Run amplitude amplification to amplify any component of $|v_{f}\rangle$ that is orthogonal to $S$.  We
run amplitude amplification for order $\epsilon^{-1/3}$ steps, and measure $P_{S}$.  Repeat this procedure
order $\epsilon^{-1/3}$ times.  If we obtain $0$ as a measurement result for any of the measurements, we
say the function is $\epsilon$-far from symmetric.
\item If the function has not been declared $\epsilon$-far from symmetric by either of the previous steps, we declare the function symmetric.
\end{itemize}
Note that a symmetric function will always be declared symmetric by this algorithm.

In order to show how the algorithm works when the function is $\epsilon$-far from symmetric,
we first need to determine how large a component orthogonal to $S$ the vector $|v_{f}\rangle$ 
will have if $f$ is $\epsilon$-far from being symmetric.  We begin by expressing the vector
$|v_{f}\rangle$ as $|v_{f}\rangle = |v_{fS}\rangle + |v_{f\perp}\rangle$, where $|v_{fS}\rangle = P_{S}|v_{f}\rangle$,
and $|v_{f\perp}\rangle = (I-P_{S})|v_{f}\rangle$.  We next define the vector
$|u_{m}\rangle$, for $m=0,1, \ldots n$, which is the superposition, with equal coefficients, of all vectors in the
computational basis with $m$ ones, e.g.
\begin{eqnarray}
|u_{0}\rangle & = & |00\ldots 0\rangle \nonumber \\
|u_{1}\rangle & = & \frac{1}{\sqrt{n}}(|100\ldots 0\rangle + |010\ldots 0\rangle + \ldots  \nonumber \\
 & & +|00\ldots 01\rangle ) .
\end{eqnarray}
We then have that 
\begin{equation}
P_{S} = \sum_{m=0}^{n} |u_{m}\rangle\langle u_{m}| .
\end{equation}
Now suppose that for the sequences with Hamming weight $m$, $f(x)=1$ for $l_{m}$ of them and $f(x)=0$
for the remaining sequences.  This implies that
\begin{equation}
\langle u_{m}|v_{f} \rangle = \frac{1}{\sqrt{N}}\left(\begin{array}{c} n \\ m \end{array}\right)^{-1/2} \left[ \left(
\begin{array}{c} n \\ m \end{array}\right) -2l_{m} \right]  ,
\end{equation}
so that
\begin{equation}
\| v_{fS}\|^{2} = 
\langle v_{f}|P_{S}| v_{f}\rangle = \sum_{m=0}^{n} \frac{1}{N}\left(\begin{array}{c} n \\ m \end{array}\right)^{-1} 
\left[ \left(\begin{array}{c} n \\ m \end{array}\right) -2l_{m} \right]^{2} .  
\end{equation} 
Next, it is relatively simple to construct the symmetric function that is closest to $f$, which we shall
call $g$.  If $x$ has Hamming weight $m$, we set $g(x)=0$ if 
\begin{equation}
l_{m}\leq \frac{1}{2}  \left(\begin{array}{c} n \\ m \end{array}\right) ,
\end{equation}
and $g(x)=1$ otherwise.  This implies that
\begin{equation}
\langle v_{f}|v_{g}\rangle = \frac{1}{N}\sum_{m=0}^{n} \left| \left(\begin{array}{c} n \\ m \end{array}\right) 
-2l_{m} \right| ,
\end{equation}
and, if $f$ is $\epsilon$-far from being symmetric, we have that $\langle v_{f}|v_{g}\rangle
< 1-2\epsilon$.  Therefore, making use of the fact that 
\begin{equation}
\left(\begin{array}{c} n \\ m \end{array}\right)^{-1} \left| \left(\begin{array}{c} n \\ m \end{array}\right) 
-2l_{m} \right| \leq 1 ,
\end{equation}
we have that 
\begin{equation}
\sum_{m=0}^{n} \frac{1}{N}\left(\begin{array}{c} n \\ m \end{array}\right)^{-1} 
\left[ \left(\begin{array}{c} n \\ m \end{array}\right) -2l_{m} \right]^{2} < 1-2\epsilon .
\end{equation}
This gives us that $\| v_{fS}\|^{2} < 1-2\epsilon$ so that $\| v_{f\perp}\|^{2} \geq 2\epsilon$.  Therefore, if $f$ is
$\epsilon$-far from being symmetric, $|v_{f}\rangle$ has a component of norm greater than or equal
to $\sqrt{2\epsilon}$ orthogonal to $S$.

Now let us look at our algorithm in more detail.  We first measure $P_{S}$ $m$ times.  If the result of any of our
measurements is $0$, we reject, and say that $f$ is not symmetric.  We will again break up our analysis into two parts.  Let $\mu = \| v_{fS}\|$ and $\mu_{0}=1-\epsilon^{2/3}$, then we will consider the two cases, $\mu < \mu_{0}$ and $\mu_{0}\leq \mu \leq (1-2\epsilon )^{1/2}$.  The second case will give us a nonzero range for $\mu$ if $\epsilon < 1/8$, which we will assume to be the case (in fact, we shall assume that $\epsilon\leq 10^{-3}$ as in the previous section).  Now, if $\mu <\mu_{0}$, the probability that $f$ passes this part of the test is $p_{m}$, where
\begin{equation}
p_{m} \leq |\mu_{0}|^{2m} = (1-\epsilon^{2/3} )^{2m} \leq e^{-2m\epsilon^{2/3}} .
\end{equation}
If we choose $m>\ln 3/(2\epsilon^{2/3})$, then this probability will be less than $1/3$.  This part of the algorithm 
requires of the order of $\epsilon^{-2/3}$ oracle calls.

Now let us look at the case when $\mu_{0}\leq \mu \leq (1-2\epsilon )^{1/2}$.
If the function has passed the first part of the test, we proceed to the second part, which makes use of the Grover
algorithm.  The Grover operator in this case is
\begin{equation}
G=(I-2|v_{f}\rangle\langle v_{f}|) (I-2P_{S}) ,
\end{equation}
and it requires two applications of the oracle to implement.  We want to analyze what happens when we 
apply this operator to $|v_{f}\rangle$, and in order to do so we define the unit vectors 
\begin{eqnarray}
|u_{1}\rangle & = & \frac{1}{\| v_{fS}\|} |v_{fS}\rangle \nonumber \\
|u_{2}\rangle & = & \frac{1}{\| v_{f\perp}\|} |v_{f\perp}\rangle .
\end{eqnarray}
The operator $G$ maps the two-dimensional space spanned by $|u_{1}\rangle$ and $|u_{2}\rangle$ into itself, 
and in the $\{ |u_{1}\rangle , |u_{2}\rangle\}$ basis, it can be represented as the $2\times 2$ matrix
\begin{equation}
G= \left( \begin{array}{cc} 2\mu^{2}-1 & 2\mu (1-\mu^{2})^{1/2} \\ -2\mu (1-\mu^{2})^{1/2} & 2\mu^{2}-1 \end{array}
\right) ,
\end{equation}
where we have set $\mu = \| v_{fS}\|<(1-2\epsilon )^{1/2}$.  The eigenvalues of this matrix are
\begin{equation}
\lambda_{\pm} = 2\mu^{2}-1 \pm 2i\mu (1-\mu^{2})^{1/2} ,
\end{equation}
with the corresponding eigenvectors given by
\begin{equation}
|\eta_{\pm}\rangle = \frac{1}{\sqrt{2}} \left( \begin{array}{c} 1 \\ \pm i\end{array} \right) .
\end{equation}
We can now calculate $G^{n}|v_{f}\rangle$. Setting 
\begin{equation}
\cos\theta = 2\mu^{2}-1\blue{,} \hspace{5mm} \sin\theta = 2\mu (1-\mu^{2})^{1/2}  ,
\end{equation}
which implies that for $\epsilon \ll 1$ that we also have $0 <\theta \ll 1$, we find that
\begin{eqnarray}
G^{n}|v_{f}\rangle & = & [ \mu \cos n\theta +(1-\mu^{2})^{1/2}\sin n\theta ] |u_{1}\rangle \nonumber \\
 & & +[-\mu \sin n\theta +(1-\mu^{2})^{1/2} \cos n\theta ] |u_{2}\rangle .
\end{eqnarray}
If we now measure $P_{S}$ in this state, the probability that we obtain one, $p(n,\mu )$, is given by
\begin{equation}
p(n,\mu ) = \cos^{2} \left[ (n-\frac{1}{2})\theta \right]  ,
\end{equation}
and the probability that we obtain $0$, $q(n,\mu )$, which would show that the function is not symmetric, is
\begin{equation}
q(n,\mu )=1-p(n,\mu ) = \frac{1}{2}\left( 1-\cos [(2n-1)\theta ] \right) .
\end{equation}

We now need to get an estimate of $\theta$.  As in the previous section, we shall assume that $\epsilon \leq 
10^{-3}$.  Note that for $(1/\sqrt{2}) \leq \mu \leq 1$, which will be true if 
$\epsilon^{2/3} < 1-2^{-1/2}$, the function $2\mu(1-\mu^{2})^{1/2}$ is monotonically decreasing.  In this case,
we have that 
\begin{equation}
2\sqrt{2} (1-\epsilon^{2/3}) \epsilon^{1/3} \geq \sin \theta \geq 2(1-2\epsilon )^{1/2} \sqrt{2\epsilon} .
\end{equation}
This implies, using the same inequality as in the last section (for $\epsilon \leq 10^{-3}$ we do have that $\theta
\leq \pi /6$), that 
\begin{equation}
\frac{2\pi \sqrt{2}}{3}(1-\epsilon^{2/3})\epsilon^{1/3} \geq \theta \geq 2\sqrt{2} (1-2\epsilon )^{1/2}\sqrt{\epsilon} ,
\end{equation}
which, for $\epsilon \leq 10^{-3}$, can be simplified to
\begin{equation}
\frac{2\pi \sqrt{2}}{3} \epsilon^{1/3} \geq \theta \geq  2\sqrt{2}(0.998)\epsilon^{1/2}  .
\end{equation}

Next, we apply $G$ $n$ times where we now choose $n$ so that $2n-1$ is the closest odd integer to
$[3/(2\sqrt{2})]\epsilon^{-1/3}$, and measure $P_{S}$.  We repeat this procedure $l$ times, where $l$ is of order $\epsilon^{-1/3}$.  If $\theta$ is near the top of its range, the probability that we will obtain $0$ when we measure $P_{S}$ is then of order one, so that our function will be shown not to be symmetric with high probability after a small number of runs.  Now let us see what happens if $\theta$ is at the bottom of its range, the worst case.  We first note that, making use of Eq.\ (\ref{estimate}), we have
\begin{equation}
\frac{1}{2\pi} [(2n-1)\theta ]^{2} \leq q(n,\mu ) \leq \frac{1}{4} [(2n-1)\theta ]^{2} .
\end{equation}
Putting in the value of $\theta$ at the bottom of its range and the value of $n$ given above gives us
\begin{equation}
\frac{1}{2\pi}(\beta \epsilon^{1/6}-2\sqrt{2}\epsilon^{1/2})^{2} \leq q(n, \mu ) \leq \frac{1}{4} \frac{1}{2\pi}
(\beta \epsilon^{1/6}+2\sqrt{2}\epsilon^{1/2})^{2} ,
\end{equation}
where $\beta = 3(0.998)$ and the order $\epsilon^{1/2}$ terms result from the fact that $2n-1$ must be an odd integer.  These are less than $1/10$ of the dominant $\beta\epsilon^{1/6}$ contribution when $\epsilon = 10^{-3}$ and decrease as $\epsilon^{1/3}$ as $\epsilon$ goes to zero.  We shall neglect them for the rest of the calcluation.  Now the probability that we will get $1$ each time we measure $P_{S}$ is
\begin{equation}
[1-q(n,\mu ) ]^{l} \leq \left( 1- \frac{1}{2\pi} \beta^{2} \epsilon^{1/3}\right)^{l} \le
 \exp \left[ -\left( \frac{l\beta^{2}}{2\pi} \right) \epsilon^{1/3}\right] .
\end{equation}
Therefore, if we choose $l> 2\pi \ln 3/(\beta^{2}\epsilon^{1/3})$, this probability can be made less than $1/3$.  The total number of oracle calls in the second part of the algorithm, that is, the part using the Grover algorithm, is of order $\epsilon^{-2/3}$, so that the entire algorithm uses order $\epsilon^{-2/3}$ oracle calls to determine whether a function is symmetric, or whether it is $\epsilon$-far from symmetric, with a probability of error of less than $1/3$. 

\section{Conclusion}
We have presented two algorithms for function property testing.  The first tells you whether a Boolean function is linear or $\epsilon$-far from linear, and if it is linear it tells you which linear function it is.  The second tells you whether a Boolean function is symmetric or $\epsilon$-far from being symmetric. Both algorithms use of the order of $\epsilon^{-2/3}$ oracle calls, independent of the number of input variables to the Boolean function.

It will be interesting to see whether quantum algorithms can be found that test for other properties of Boolean functions.  The Bernstein-Vazirani algorithm and amplitude amplification give us a powerful tools, which are not available in the classical case.  It remains to be seen exactly how useful they can be.

\section*{Acknowledgments}
One of us (MH) was supported by the National Science Foundation under grant PHY-0903660 and by a PSC-CUNY grant. EA acknowledges partial support from EPSRC grant EP/G009821/1.

\end{document}